\documentclass[12pt]{article}
\usepackage{enumitem}
\usepackage[english]{babel}
\usepackage[utf8]{inputenc}
\usepackage[authoryear,comma]{natbib}
\usepackage{amsmath,multibib}
\usepackage{bbm}
 \usepackage{relsize}
\usepackage{amssymb,amsthm,mathabx,mathtools}
\usepackage{mathrsfs}
\usepackage{graphicx}
\usepackage{url}
\usepackage{caption,subcaption}
\usepackage{booktabs}
\usepackage{multirow}
\usepackage{caption}
\usepackage{epsfig,float}
\usepackage{url} 
\usepackage[usenames,dvipsnames]{color} 
\usepackage{listings}
\usepackage{mathrsfs,bm} 

\newcommand*\circled[1]{\tikz[baseline=(char.base)]{
            \node[shape=circle,draw,inner sep=2pt] (char) {#1};}}

\usepackage{graphicx}

\theoremstyle{plain}
\newtheorem{thm}{Theorem}

\newcommand{\bthm}{\begin{thm}}
\newcommand{\ethm}{\end{thm}}
\newcommand{\bpf}{\begin{proof}}
\newcommand{\epf}{\end{proof}}
\theoremstyle{definition}
\newtheorem{defn}{Definition}

\newtheorem{rem}{Remark}

\usepackage{deep-macro}
\usepackage{epigraph}

\usepackage{mathtools}

\usepackage{epigraph}
\usepackage{framed}
\usepackage{relsize}
\usepackage[most]{tcolorbox}
\usepackage{tikz}
\usetikzlibrary{automata, positioning}
\usetikzlibrary{arrows, arrows.meta, calc, positioning, quotes, shapes, shapes.geometric}
\tikzstyle{block} = [rectangle, draw, text centered,    text width=7em, rounded corners, minimum height=1cm]
\tikzstyle{line} = [draw, -latex',line width=.22mm]

\usepackage{bbm}
\usepackage{tabularx}
\usepackage{tabularx}
\newcolumntype{Y}{>{\centering\arraybackslash}X}
\newcolumntype{f}{>{\centering\arraybackslash}X}
\newcolumntype{h}{>{\hsize=.5\hsize\centering\arraybackslash\extracolsep{.1em}}X}

\newcolumntype{C}[1]{>{\hsize=#1\hsize\centering\arraybackslash}X}

\usepackage{deep-macro}
\newcommand{\dFnull}{d \hspace{-.08em}\circ \hspace{-.08em}F_0}

\usepackage[bottom]{footmisc}
\usepackage{cancel}

\usepackage{JASA_manu}
\usepackage{setspace}

\begin{document}

\title{
\vspace{-3em}
Abductive Inference and C.~S.~Peirce:  150 Years Later
}
\author{
\centering
\begin{tabularx}{.97\linewidth}{ff}
Subhadeep (DEEP) Mukhopadhyay\\
Email: \texttt{deep@unitedstatalgo.com}
\end{tabularx} }

\date{}
\maketitle





\begin{abstract} 
{\small \textit{Two pillars of the paper}. This paper is about two things: (i) Charles Sanders Peirce (1837–1914)---an iconoclastic philosopher and polymath who is among the greatest of American minds. (ii) Abductive inference---a term coined by C. S. Peirce, which he defined as ``the process of forming explanatory hypotheses. It is the only logical operation which introduces any new idea.''

\vskip.25em
\textit{Abductive inference and quantitative economics}. Abductive inference plays a fundamental role in empirical scientific research as a tool for discovery and data analysis. Heckman and Singer (2017) strongly advocated ``Economists should \textit{abduct}.'' Arnold Zellner (2007)  stressed that ``much \textit{greater} emphasis on reductive [abductive] inference in teaching econometrics, statistics, and economics would be desirable.''  But currently, there are no established theory or practical tools that can allow an empirical analyst to abduct. This paper attempts to fill this gap by introducing new principles and concrete procedures to the Economics and Statistics community. I termed the proposed approach as Abductive Inference Machine (\texttt{AIM}).
\vskip.15em
\textit{The historical Peirce's experiment}. In 1872, Peirce conducted a series of experiments to determine the distribution of response times to an auditory stimulus, which is widely regarded as one of the most significant statistical investigations in the history of nineteenth-century American mathematical research \citep{stigler1978}. On the 150th anniversary of this historical experiment, we look back at the Peircean-style abductive inference through a modern statistical lens. Using Peirce’s data, it is shown how empirical analysts \textit{can abduct} in a systematic and automated manner using \texttt{AIM}.}

\end{abstract} 






\newpage 

\section{Introduction}

Charles Sanders Peirce (1839–1914), America's greatest philosopher of science, was also a brilliant statistician and experimental scientist. For 32 years, from 1859 until 1891, he worked for the United States Coast and Geodetic Survey\footnote{U. S. Coast and Geodetic Survey was established on February 10, 1807, by President Thomas Jefferson. It was the nation’s first civilian scientific agency.}. During this time, he developed an unfailing passion for experimental research. He was deeply involved in developing theoretical and practical methods for acquiring high-precision scientific measurements, which ultimately earned him an international reputation as an expert in `measurement error' in physics. Robert \cite{crease2009charles}, a philosopher and historian of science, noted:  `His [Peirce's] work helped remove American metrology from under the British shadow and usher in an American tradition.'

\begin{figure}[h]
\vskip.4em
  \centering
\includegraphics[width=.75\linewidth,keepaspectratio,trim=1.8cm 1cm 1.8cm 1cm]{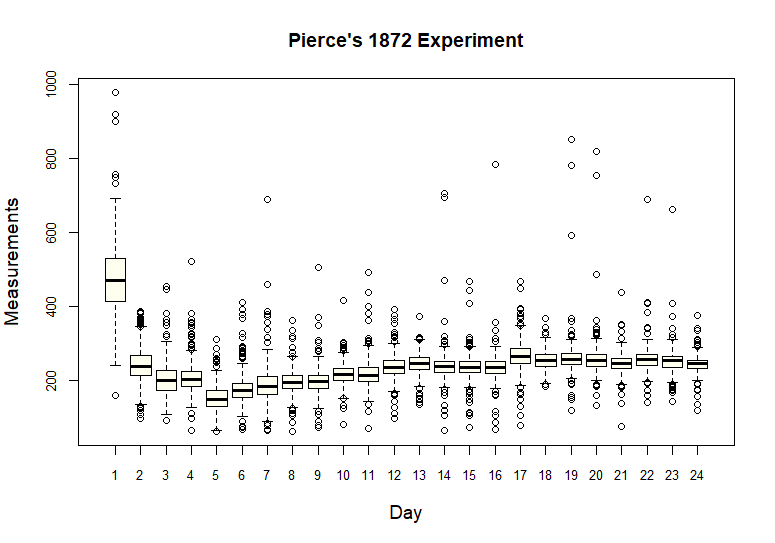}
\vskip.3em
\caption{First look at Peirce's auditory response data. 
The x-axis denotes 24 different days of experiments and the y-axis displays the data of individual experiment as boxplot.}\label{fig:boxplot}
\end{figure}
\textit{1872 Experimental Data}. In 1872, he conducted a series of famous experiments to determine the distribution of response times to an auditory stimulus. He measured the time that elapsed between the making of a sharp sound and the record of reception of the sound by an observer, employing a Hipp chronoscope (some kind of sophisticated clock). Fig. \ref{fig:boxplot} shows the dataset, which consists of roughly $500$ measurements (recorded in nearest milliseconds) each day for $k=24$ different days\footnote{For further details on the experimental setup and the full dataset, consult the online Peirce Edition Project: vol 3, p. 133--160 of the chronological edition \citep{peirce2009vol3}. It's also available in the R-package \texttt{quantreg}. }. Note that the first-day observations are systematically different from others (also called systematic bias), and the inconsistency was due to the lack of experience of the observer, which was corrected on the next day. The next 23 days show much more consistent (comparable) measurements.








\section{Gauss' Law of Error}
\begin{quote}
    \textit{What observation has to teach us is [density] \textit{function}, not a mere number.}    
\begin{flushright}
\vspace{-.36em}
{\rm --- C. S. \cite{peirce1873}} \end{flushright}
\end{quote}
\textit{Deciphering the Law of Errors}. Peirce's actual motivation for doing the experiment was to study the probabilistic laws of fluctuations (also called errors) in the measurements and to investigate how response time distributions differ from the standard Gauss' law.
\vskip.7em

{\bf Nineteenth-century statistical learning}. \cite{peirce1873} presented a detailed empirical investigation of the reaction-time densities for each day. He was driven by two goals: to understand the shape of the reaction time densities and to compare them with the expected Gaussian distribution. His approach had a remarkably modern conceptual basis: first, he developed smooth kernel density-type probability density estimates to understand the shape of error distributions\footnote{Peirce made a pioneering contribution to American statistics by developing the concepts that underpin nonparametric density estimation.};
second, he performed a goodness-of-fit (GOF) type assessment through visual comparison between the shape of Gaussian distribution and the nonparametrically estimated densities\footnote{
However, at that time no theory of GOF was available. It took 30 more years for an English mathematician, Karl Pearson, to make the breakthrough contribution in developing the formal language of the GOF.}, and concluded that the reaction-time distributions `differed very little from' the expected normal probability law\footnote{The term ``normal distribution'' was coined by Peirce.}.  



\vskip.75em
{\bf Twentieth-century statistical learning}.  Almost sixty years later, the same dataset was reanalyzed by \cite{wilson1929}, and they came to a very different conclusion. Wilson and Hilferty performed a battery of tests to verify the appropriateness of the normal distribution. For each series of measurements, they computed $23$ statistics (e.g., mean, standard deviation, skewness, kurtosis,  interquartile range, etc.) to justify substantial departures from Gaussianity.  Interestingly, without any formal statistical test, simply by carefully looking at the boxplots in Fig. \ref{fig:boxplot}, we can see the presence of significant skewness (the median cuts the boxes into two unequal pieces), heavy-tailedness (long whiskers relative to the box length), and ample outlying observations---which is good enough to suspect the adequacy of Gaussian distribution as a model for the data. 

\begin{rem}
The non-Gaussian nature of the error distribution of scientific measurements is hardly surprising\footnote{Even \cite{wilson1929} noted the same: `according to our previous experience such long
series of observations generally reveal marked departures from the normal law.' }--- in fact, it is the norm, not the exception \citep{bailey2017not}, which arises primarily because it is hard to control all the factors of a complex measurement process. But what is startling is that even Peirce's experiment, a simple investigation of recording response times with the same instrument by the same person under more or less similar conditions, can produce so much heterogeneity.
\end{rem}




\begin{quote}
    \small{`\textit{Does statistics help in the search for an alternative hypothesis? There is no codified statistical methodology for this purpose. Text books on statistics do not discuss either in general terms or through examples how to elicit clues from data to formulate an alternative hypothesis or theory when a given hypothesis is rejected.}'}
   \begin{flushright}
 \vspace{-.4em}
 {\rm --- C. R. \cite{rao2001statistics}} \end{flushright}
\end{quote}
{\bf Revised Goal: From Testing to Discovery}. Confirmatory analysis through hypothesis testing provides investigators absolutely no clues on what might actually be going on.\footnote{An average statistician uses data to confirm or reject a particular theory/model. A competent statistician uses data to \textit{sharpen} their theory/model.} Simply rejecting a hypothesis---saying that it is non-Gaussian---does not add any new insight into the underlying laws of error. Thus, our focus will be on developing a data analysis technique that can identify the most questionable aspects of the existing model and can \textit{also provide} concrete recommendations on how to rectify those deficiencies in order to build a better and more realistic model for the measurement uncertainties.



\section{The Problem of Surprise}\label{sec:surp}
\begin{quote}
    \textit{It is not enough to, look for what we anticipate. The greatest gains from data come from surprises. \hfill ---  John \cite{tukey1972exploratory}}    
\end{quote}



{\bf Modeling the Surprise}. All empirical laws are approximations of reality---sometimes good, sometimes bad. We will be fooling ourselves if we think there is a single best model that fits Peirce's experimental data. Any statistical model, irrespective of how sophisticated it is, should be ready to be surprised by data. The goal of empirical modeling is to develop a general strategy for describing how a model should react and adapt itself to reduce the surprise.


\vskip.5em
Without surprise, there is no discovery. The `process' of discovering \textit{new} knowledge from data starts by answering the following questions: Is there anything surprising in the data? If so, what  makes it surprising? How should the current model react to the surprise? How can it modify itself to rationalize the empirical surprise? To develop a model and principle for statistical discovery,  we need to first address these fundamental data modeling questions. In subsequent sections, we develop one such general theory.
\vskip.5em




Basic notations used throughout the paper: $X$ is a continuous random variable with cdf $F(x)$, pdf $f(x)$. The quantile function is given by $Q(u)=F^{-1}(u)$. Expectation with respect to the initial working model $F_0(x)$ is abbreviated as $\Ex_0(\psi(X)) :=\int \psi \dd F_0$, and expectation with respect to the empirical $\wtF$ is simply written as $\wtE(\psi(X)) :=\int \psi \dd \wtF$. The inner product of two functions $\psi_1$ and $\psi_2$ in $\cL^2(dF_0)$ will be denoted by $\langle \psi_1, \psi_2 \rangle_{F_0}:=\int \psi_1 \psi_2 \dd F_0$.

\section{A Model for Empirical Discovery}
\label{sec:theory}
\begin{quote}
    \textit{There is no established practice for dealing with surprise,
even though surprise is an everyday occurrence. Is there a best way to respond to empirical surprises?}    \begin{flushright}
\vspace{-.24em}
{\rm ---   \cite{heckman2017abducting}} \end{flushright}
\vspace{-.74em}
\end{quote}
\subsection{A Dyadic Meta-Model} 
Science is a ``self-corrective'' enterprise that seeks new knowledge by refining the known.\footnote{According to Peirce, every branch of scientific inquiry exhibits ``the vital power of self-correction'' that permits us to make progress and grow our knowledge; see, \cite{stan-peirce}.} The same is true for statistical modeling: it explores and discovers unknown patterns by smartly utilizing the known (expected) model.  In the following, we formalize this general principle.

\begin{defn}[A Dyadic Meta-Model] \label{def1:2tm}
$X$ be a continuous random variable with true unknown density $f(x)$. Let $f_0(x)$ represents a possibly misspecified predesignated working model for $X$, whose support includes the support of $f(x)$. Then the following density decomposition formula holds:

\beq \label{eq:fgd}
f(x)\,=\,f_0(x)\,d\big(F_0(x);F_0,F\big), \eeq
where the $d(u;F_0,F)$ is defined as 
\beq d(u;F_0,F)= \dfrac{f(F_0^{-1}(u))}{f_0(F_0^{-1}(u))}, ~\,0<u<1,\eeq
which is called `comparison density' because it \textit{compares} the initial model-0 $f_0(x)$ with the true $f(x)$ and it integrates to one:
\[\int _0^1 d(u;F_0,F)\dd u \,=\, \int_x d(F_0(x);F_0,F) \dd F_0(x) \,=\,\int_x \big(f(x)/f_0(x)\big) \dd F_0(x)\,=\, 1. ~~\]
To simplify the notation, $d(F_0(x);F_0,F)$  will be abbreviated as $d_0(x)$. One can view \eqref{eq:fgd} as a ``meta-model''---a model comprising two sub-models that blends existing imprecise knowledge $f_0(x)$ with new empirical knowledge $d_0(x)$ to provide us complete picture of the uncertainty. 



\end{defn}
The above density representation formula can be interpreted from many different angles: 
\vskip.55em
1. {\bf Model-Editing Tool}. The dyadic model provides a general statistical mechanism for designing a ``better'' model by editing  or sharpening the existing version. For that reason, we call $d$ the density-sharpening function (DSF). Next section presents how to learn \texttt{DSF} from data. The $d$-modulated repaired $f_0$-density in Eq. \eqref{eq:fgd} will be referred to $d$-sharp $f_0$. 

\vskip.55em

2. {\bf Surprisal Function}. The process of data-driven discovery starts with a surprise---a deviation between the data and the expected model. The \texttt{DSF} $d(u;F_0,F)$ gets activated only when model-0 encounters surprise, and its shape encodes the \textit{nature} of surprise. When there is no surprise, $d(u;F_0,F)$ takes the shape of a `flat' uniform density.


\textit{Surprise to information gain}: 
It is not enough to simply detect an empirical surprise. For statistical learning, it is critical to know: What information can we gain from the observed surprise? And how can we use that information to revise our initial model of reality?  The density-sharpening function $d(u;F_0,F)$ provides a pathway from surprise to information gain that bridges the gap between the initial belief and knowledge.

\vskip.55em

3.  {\bf Simon's Means-Ends Analysis}. The model \eqref{eq:fgd} interacts with the outer data environment through two information channels: 
\begin{itemize}[itemsep=2pt,topsep=1.6pt]
    \item Afferent (or `inward') information: it captures and represents the `\textit{difference}' between the desired and present model using $d_0(x)$\footnote{
It would be pointless to waste computational resources on the redundant part of the data.}.

    \item Efferent (or `outward') information: it intelligently searches and provides the best course of `\textit{actions}' that changes the present model through \eqref{eq:fgd} to reduce the difference\footnote{$d_0(x)$ ``fires'' \textit{actions} when the \textit{difference} in information content between $F_0$ and $\wtF$ reaches a threshold.}.
\end{itemize}
Herbert \cite{simon1988science} noted that \textit{any} general-purpose computational learning system must have these two information processing components. Models equipped with this special structure are known as the `Means-Ends analysis model' in the artificial intelligence community.

\vskip.55em

4. {\bf Detective's Microscope}\footnote{The name was inspired from John \citet[p. 52]{tukey1977}}. Information in the data can be broken down into two parts:
\vspace{-.6em}
\beq
\text{Data Information ~ = ~\,Anticipated part ~+ ~Unexpected surprising part.}~~~
\vspace{-.6em}
\eeq 
Model-0 explains the first part, whereas $d_0(x)$ captures everything that is unexplainable by the initial $f_0(x)$. Accordingly, $d_0(x)$ performs dual tasks: it reveals the incompleteness of our starting assumptions and provides strategies on how to revise it to account for the observed puzzling facts. In short,  $d(u;F_0,F)$ plays the role of a  \textit{detective's microscope}, permitting data investigators to assemble clues to initiate a systematic search for new explanatory hypotheses that fit the evidence and solve the puzzle.



\vskip.55em

5.  {\bf System-1 and System-2 Architecture}\footnote{This `Two Systems' analogy was inspired by Daniel Kahneman's work on `Thinking, Fast and Slow.'}.
In our dyadic model, System-1 is denoted by $f_0(x)$ that captures the background knowledge component. Model-0 interacts with the environment through System 2 $d$-function. The DSF $d$ allows model-0 to self-examine its limitations and also offers strategies for self-correcting to adapt to new situations.  The \texttt{DSF} plays the role of a `supervisor' whose goal is model management. It helps the subordinate $f_0$ to figure out what's missing and how to fix this. Our dyadic model combines both system-1 and system-2 into one integrated modeling system.




\vskip.55em

6. {\bf A Change Agent}: The great philosopher Heraclitus taught us that change is the only constant thing in this world. If we believe in this doctrine then we should focus on modeling the change, not the model itself.\footnote{Isaac Newton confronted a similar problem in the mid-1600s: He wanted to describe a falling object, which changes its speed every second. The challenge was: How to describe a ``moving'' object? His revolutionary idea was to focus on modeling the ``change,'' which led to the development of Calculus and Laws of Motion. Here we are concerned with a similar question: How to change probability distribution when confronted with new data? In our dyadic model \eqref{eq:fgd}, the sharpening function $d$ provides the necessary ``push to change.''} The dyadic model operationalizes this philosophy by providing a universal law of model evolution: how to produce a useful model by changing an imperfect model-0 in a data-adaptive manner. The rectified $f_0(x)$ inherits new characteristics through $d_0(x)$ that give them a better chance of survival in the new data environment.






\subsection{A Robust Nonparametric Estimation Method} \label{sec:rtheory}
To operationalize the density-sharpening law, we need to estimate from data the function $d_0(x)$, which is the \textit{cause of change} in the state of a probability distribution. We describe a theory of robust nonparametric estimation whose core concepts and methodological tools are introduced in a `programmatic' style---making it easy to translate the theory into a concrete algorithm.

\begin{rem}[What are we approximating?]
Before describing the approximation theory, it is vital to emphasize that, unlike traditional nonparametric (or machine learning) methods where the goal is to produce a density estimate $\hf(x)$, in our case, the focus is on estimating the sharpening kernel $\whd_0(x)$---the `gap' between theory and measurements. This will provide \textit{rational} explanations for the surprising facts and, because of Eq. \eqref{eq:fgd}, concurrently rectify the initial model $f_0$. Also, see sec. \ref{sec:AIMsec}.
\end{rem}


{\bf Step 0. Data and Setup}. We observe a random sample $X_1,\ldots,X_n \mathrel{\dot\sim} F_0$. By ``$\mathrel{\dot\sim}$'' we mean 
that $F_0$ is a tentative (approximate theoretical) model for $X$ that is given to us. And, $f(x)$ denotes the unknown true model from which the data were generated.

\vskip.75em

{\bf Step 1. LP-orthogonal System}. Note that \texttt{DSF} $d_0(x)$ is a function of rank-$F_0$ transform (i.e.,  probability integral transform with respect to the base measure) $F_0(X)$. Hence, one can \textit{efficiently} approximate $\dFnull(x) \in \cL^2({dF_0})$ by expanding it in a Fourier series of polynomials that are function of $F_0(x)$ and orthonormal with respect to the user-selected base-model $f_0(x)$. One such orthonormal system is the LP-family of rank-polynomials \citep{deep18nature, Deep17LPMode}, whose construction is given below.

LP-basis construction for an arbitrary continuous $F_0$: Define the first-order LP-basis function as \textit{standardized} rank-$F_0$ transform:
\beq \label{eq:T1} 
 T_1(x;F_0)\,=\,\sqrt{12} \big\{F_0(x) - 1/2\big\}. \eeq
Note that $\Ex_0(T_1(X;F_0))=0$ and $\Var_0(T_1(X;F_0))=1$. Next, apply Gram-Schmidt procedure on $\{T_1^2, T_1^3,\ldots\}$ to construct a higher-order LP orthogonal system $T_j(x;F_0)$:
\vspace{-1em}
\bea 
T_2(x;F_0) &=& \sqrt{5} \big\{ 6 F^2_0(x)  - 6 F_0(x) + 1 \big\}   \\
T_3(x;F_0) &=&  \sqrt{7} \big\{  20 F^3_0(x)  - 30F^2_0(x)  + 12F_0(x)     -1   \big\}   \\
T_4(x;F_0) &=&       \sqrt{9} \big\{ 70F^4_0(x)  - 140F^3_0(x)  + 90F^2_0(x)  -20F_0(x)     +1   \big\},
\eea
and so on. For data analysis, we compute them by executing the Gram-Schmidt process numerically. Hence, there is no need for bookkeeping the explicit formulae of these polynomials. By construction, the LP-sequence of polynomials satisfy the following conditions: 
\beq \label{eq:lponp}
\int\nolimits_x  T_j(x;F_0) \dd F_0\,=\,0
;~~\,\int\nolimits_x  T_j(x;F_0)T_k(x;F_0)\dd F_0=\delta_{jk},
\eeq
where $\delta_{jk}$ is the Kronecker’s delta function. 
The notation for LP-polynomials $\{T_j(x;F_0)\}$ is meant to emphasize: (i) they are polynomials of $F_0(x)$ (not raw $x$) and hence are inherently robust. (ii) they are orthonormal with respect to the distribution $F_0$, since they satisfy \eqref{eq:lponp}. We also define the Unit LP-bases via quantile transform: $S_j(u;F_0)=T_j(Q_0(u);F_0)$, $0\le u \le 1$.

\vskip.77em

{\bf Step 2. LP-Fourier Approximation}. LP-orthogonal series representation of the density-sharpening function $d_0(x)$ is given by 
\beq \label{eq:l2d}
d_0(x) := d(F_0(x);F_0,F)\,=\,1+\sum_j \LP[j;F_0,F] \,T_j(x;F_0),
\eeq 
where the expansion coefficients $\LP[j;F_0,F]$ satisfy
\beq \label{eq:lpcoef}
\LP[j;F_0,F]= \big\langle \dFnull, T_j \big \rangle_{F_0},~~~(j=1,2,\ldots,).
\eeq

{\bf Step 3. Nonparametric Estimation}. To estimate the LP-Fourier coefficients from data, rewrite Eq. \eqref{eq:lpcoef} in the following form:  
\beq 
\LP[j;F_0,F]= \int_{-\infty}^{\infty} d_0(x) T_j(x;F_0) f_0(x) \dd x = \int_{-\infty}^{\infty} T_j(x;F_0) f(x) \dd x = \Ex_F[T_j(X;F_0)]
\eeq
which expresses $\LP[j;F_0,F]$ as the expected value of $T_j(X;F_0)$. Accordingly, estimate the LP-parameter as

\beq \label{eq:lpest}
\tLP_j := \LP[j;F_0, \wtF] \,=\, \wtE[T_j(X;F_0)] \,=\, \frac{1}{n} \sum_{i=1}^n T_j(x_i;F_0).
\eeq
These expansion coefficients act as the coordinates of true $f(x)$ \textit{relative to} assumed $f_0(x)$:
\beq \label{eq:corLP}
\big[ F \big]_{F_0}~ :=~ \Big(\LP[1;F_0,\wtF], \ldots, \LP[m;F_0,\wtF]\Big),~~\,1\le m < r~~~
\eeq
where $r$ is the number of unique values observed in the data.   
\vskip.65em
{\bf Step 4. Surprisal Index}. Can we quantify the surprise of $f_0$ when it comes in contact with the new data? We define the surprise index of the hypothesized model as follows:

\beq \label{eq:sindex}
{\rm SI}(F_0, F)\,=\,\sum_{j} \Big | \LP[j; F_0, F]   \Big|^2
\eeq
which can be computed by substituting \eqref{eq:lpest} into \eqref{eq:sindex}. The motivation behind this definition comes from recognizing that ${\rm SI}(F_0, F)=\int_0^1 d^2 -1$, i.e., the divergence-measure \eqref{eq:sindex} captures the departure of $d(u;F_0,F)$ from uniformity. Note that when $d$ takes the form of ${\rm Uniform}(0,1)$, then no correction is required in \eqref{eq:fgd}---i.e., the assumed model $f_0(x)$ is capable of fully explaining the data without being surprised at anything. An additional desirable property of the measure \eqref{eq:sindex} is that it is invariant to monotone transformations of the data. 

\begin{rem}[Information is an \textit{observer-dependent} concept]
Our definition \eqref{eq:sindex} is different from the classical Shannon-style measure of surprise or information. We view surprise as a ``fundamentally relativistic,'' not an absolute quantity. The same data can have different surprising information content for different background-knowledge-based initial models/agents. In more philosophical terms: ${\rm SI}(F_0, F)$ captures observer-specific \textit{useful} information of a dataset. 
\end{rem}

{\bf Step 5. Key Elements of Surprise}. A `large' value of ${\rm SI}(F_0, F)$ indicates that the model $f_0(x)$ got shocked by the data. But what caused this? This is the same as asking: what are the main `broken components' of the initial believable model $f_0(x)$ that need repair? As George \cite{box1976science} said: 

\begin{quote}
    \textit{Since all models are wrong the scientist must be alert to what is importantly wrong. It is inappropriate to be concerned about mice when there are tigers abroad.}    
\end{quote}

Note that the value of $\tLP_j$ is expected to be ``small'' when underlying distribution is close to the assumed $F_0$; verify this from \eqref{eq:lpest}. We discuss two pruning strategies that effectively remove the noisy LP-components that can cause the density estimate $\hf(x)$ to be unnecessary wiggly.  Identify the `significant' non-zero LP-coefficients\footnote{Under the null model, sample LP-statistic follows asymptotically $\cN(0,n^{-1/2})$.} with $|\tLP_j|>2/\sqrt{n}$. One can further refine the denoising method as follows: sort them in descending order based on their magnitude (absolute value) and compute the penalized ordered sum of squares. This Ordered PENalization scheme will be referred as \texttt{OPEN}:
\beq \label{eq:open}
\texttt{OPEN}(m)~=~\text{Sum of squares of top $m$ LP coefficients}~-~\dfrac{\gamma_n}{n}m.~~~~~~\eeq
For AIC penalty choose $\gamma_n=2$, for BIC choose $\gamma_n=\log n$, etc. For more details see \cite{D20copula} and \cite{Deep17LPMode}. Find the $m$ that maximizes the ${\rm OPEN}(m)$. Store the selected indices $j$ in the set $\cJ$; the set of functions $\{T_j(x;F_0)\}_{j \in \cJ}$ then denote the key `surprising directions' that need to be incorporated into the current model to make it data-consistent. The \texttt{OPEN}-smoothed LP-coefficients will be denoted by $\widehat{\LP}_j$.


\vskip.77em
{\bf Step 6. MaxEnt Lazy Update}. We build an improved exponential density estimate for $d(u;F_0,F)$, which, unlike the previous $\sL^2$-Fourier series model \eqref{eq:l2d}, is guaranteed to be non-negative estimate and integrates to 1. The basic idea is to choose a model for $d$ to sharpen $f_0$ in order to provide a better explanation of the data by \textit{minimizing} surprises as much as possible.  We can formalize this idea using the notion of relative-entropy (or Kullback-Leibler divergence) between $f_0$ and the $d$-sharp $f_0$:
\beas 
{\rm KL}(f_0 \| f) &=&\int f(x) \log \Big\{\frac{f(x)}{f_0(x)}\Big\} \dd x~~~~~~\\
&=& \int \frac{f(x)}{f_0(x)} \log \Big\{\frac{f(x)}{f_0(x)}\Big\} \,f_0(x) \dd x~~~~~~\\
&=& \int d_0(x) \log\{d_0(x)\} \,f_0(x) \dd x~~~~~~
\eeas
Since $d_0(x)= d(F_0(x);F_0, F)$, substituting $F_0(x)=u$ in the above equation, we get the following important result in terms of entropy of $d$: $H(d)=-\int d \log d$
\beq \label{eq:KLsur}
{\rm KL}(f_0 \| f)=\int_0^1 d \log d = - {\rm Entropy}(d),
\eeq 
which can also be viewed as a measure of surprise.  Thus the goal of searching for $d$ by minimizing the KL-divergence between the old and new model reduces to the problem of finding a $d$ by maximizing its entropy. This is known as the principle of maximum entropy (MaxEnt), first expounded by E.T. \cite{jaynes1957}.\footnote{See, for example, the work of Amos \cite{golan2018book} and Esfandiar \cite{maasoumi1993} for an excellent review of the usefulness of `maxent information-theoretic thinking' for econometrics and decision sciences. Additional recent works on the application of maximum-entropy techniques in empirical economics can be found in  \cite{buansing2020, mao2020information, lee2021maximum}.} However, an maximization of $H(d)=-\int_0^1 d \log d$ under the normalization constraint $\int_0^1 d =1$, among all continuous distributions supported over unit interval, will lead to the trivial solution:
\[ d(u;F_0,F)\, = \,1, ~~0<u<1.\vspace{-.2em}\]
\begin{rem}
Despite its elegance, the classical Jaynesian inference is an incomplete data modeling principle since it only tells us how to assign probabilities, not how to design and select appropriate constraints. Discovery, by definition, can't happen by \textit{imposing preconceived} constraints. The core `intelligence' part of any empirical modeling involves appropriately designing and searching for \textit{relevant} `directions' (constraints) that neatly capture the surprising information. More discussion on this is given in \cite{D21maxentcop}.
\end{rem}
\textit{Law of Lazy Update}. The key question is: how to determine the informative constraints? Jaynes' maximum entropy principle remains completely silent on this issue and assumes we know the relevant constraints \textit{ab initio} (i.e., sufficient statistic functions)---which, in turn, puts restrictions on the possible `shape' of the probability distribution. We avoid this assumption as follows, using what we call `Law of Lazy Update': (i) Identify a small set of most important LP-sufficient statistics functions\footnote{These sets of specially-designed functions provide the simplest and most likely explanation of how the model $f_0$ differs from reality.} using Step 5, which filters out the `directions' where close  attention should be focused.  (ii) Find a sparse (smoother) probability distribution by maximizing the entropy 
$H(d)$ under the normalization constraint $\int d=1$ and the following LP-moment constraints:
\beq \label{eq:cons}
\Ex[S_j(U;F_0)]\,=\,\LP[j;F_0,\wtF],~~~\text{for $j\in \cJ$}.
\eeq
The solution of the above maxent-constrained optimization problem can be shown to take the following exponential \eqref{eq:dexp} form 
\beq \label{eq:dexp}
d_{\teb}(u;F_0,F)\,=\,\exp\Big \{ \sum_{j\in \cJ} \te_j S_j(u;F_0)\,-\, \Psi(\teb)\Big \},~~0<u<1
\eeq
where $\Psi(\teb)=\log \int_0^1 \exp\{ \sum_j \te_j S_j(u;F_0)\}\dd u.$
\begin{rem}[Economical and Explanatory Construction]
The principle of `maxent lazy update' provides a model for $d(u;F_0,F)$, which acts as a \textit{policymaker} for $f_0$---one who formulates the preferred course of action on how to amend the existing model $f_0$ (incorporating eqs \ref{eq:cons}) in a cost-effective way to achieve the most ``economical description'' of the current reality.
\end{rem}


\begin{rem}
Incidentally, Peirce also had a strong interest in building `economical' models\footnote{Also see Peirce's 1979 article on ``Economy of Research,'' which is widely regarded as the first real attempt to establish the fundamental principles of \textit{marginal utility theory}. Stephen Stigler brought this to my attention.} and was influenced by the English philosopher William of Ockham. During the 1903 Harvard Lectures on Pragmatism, Peirce noted: ``There never was a sounder logical maxim of scientific procedure than Ockham’s razor: Entia non sunt multiplicanda praeter necessitatem.''
\end{rem}
\begin{rem}[Rational agent interpretation]
$d_0(x)$ acts as a rational agent for $f_0(x)$, which designs and selects best possible actions (alternatives) by minimizing the surprise \eqref{eq:KLsur}, subject to the constraints \eqref{eq:cons}. This kind of rationalistic empirical models were called \textit{machina economicus} by \cite{parkes2015economic}.
\end{rem}

\begin{rem}
Our style of learning of $d$ function from data performs two critical operations: The formation of new hypotheses (design of LP-sufficient statistic functions of $d$) and selection or adoption of some of the most prominent ones (through \texttt{OPEN} model selection). 
\end{rem}

\vspace{-.65em}

\subsection{Repair-Friendly  {\boldmath$\DS(F_0,m)$} Models}
\label{sec:rfmodel}

\begin{defn} \label{def:ds}
$\DS(F_0,m)$ stands for {\bf D}ensity-{\bf S}harpening of $f_0(x)$ using $m$-term LP-series approximated $d_0(x)$. Two categories of $\DS(F_0,m)$ class of distributions are given below: 
\beq  \label{DSm1}
\text{Orthogonal series $\DS(F_0,m)$}:~~~~~f(x)\,=\,f_0(x)\Big[ 1\,+\, \sum_{j=1}^m \LP[j;F_0,F]\, T_j(x;F_0)\Big],~~~~~
\eeq
\vspace{-1.65em}
\beq  \label{DSm2}
\text{Maximum Entropy $\DS(F_0,m)$}:~~~~~ f(x)\,=\,f_0(x) \exp\Big \{ \sum_{j=1}^m \te_j T_j(x;F_0)\,-\, \Psi(\teb)\Big\}.~~~~~~\,
\eeq

They are obtained by replacing \eqref{eq:l2d} and \eqref{eq:dexp}, respectively, into the dyadic model \eqref{eq:fgd}. 
The truncation point $m$ indicates the search-radius around the expected $f_0(x)$ to create permissible models. $\DS(F_0,m)$ models with higher $m$ entertains alternative models of higher complexity. However, to exclude absurdly rough densities, it is advisable to  focus on the vicinity of $f_0$ by choosing an $m$ that is not too large. In our experience, $m=6$ (or at most $8$) is often sufficient for real data applications---since $f_0(x)$ is a knowledge-based sensible starting model.
\end{defn}

\begin{rem}
The goal of empirical science is to progressively sharpen the existing knowledge by discovering \textit{new} patterns in the data, thereby leading to a new revised theory. The `density-sharpening' mechanism facilitates and automates this process. 
\end{rem}

\begin{rem}[Blending the old with the new]
The above density-editing schemes modify the initial probability law $f_0(x)$
with a small set of new additional `shape functions' (i.e., LP-sufficient statistics $\{T_j(x;F_0)\}_{j\in \cJ}$) that serve as \textit{explanations} for the surprising phenomenon. This will be more clear in the next section where we carry out Peirce data analysis using the density-sharpening principle, governed by the simple general law described in Definition \ref{def:ds}.
\end{rem}

\begin{rem}
$\DS(F_0,m)$ models can be viewed as `descent with modification,' which partially inherits characteristics of model-0 and adds some new shapes to it. This shows how new models are born out of a pre-existing inexact model with some modification dictated by the density-sharpening kernel $d_0(x)$
---thereby helping $f_0(x)$ to broaden its initial knowledge repertoire. 
\end{rem}

\begin{rem}[Model Economy]
Recall, in Section \ref{sec:surp}, we raised the question: how should a model adapt and generalize in the face of surprise? 
$\DS(F_0,m)$ is a class of nonparametrically-modified parametric models that precisely answer this question. In particular, density models \eqref{DSm1} and \eqref{DSm2} allow modelers to `fix' their broken models (in a fully automated manner) rather than completely replacing them with a brand new model built from scratch. Two practical advantages of constructing auto-adaptive models: Firstly, it reduces the waste of computational resources, and secondly, it extends the life span of of the initial, imprecise knowledge-model $f_0(x)$ by making it reusable and sustainable---we call this ``model economy.''
\end{rem}


\vspace{-.69em}




\section{Laplace's Two Laws of Error} 
\label{sec:analysis}
Our search for the laws of errors begins with the question: what is the most natural choice of the error distribution that we anticipate to hold at least approximately. Two candidates are: 
\begin{itemize}[topsep=8pt,itemsep=11pt, parsep=6pt]
 \setlength{\itemindent}{-.6em}
    \item Laplace distribution. In 1774, i.e., almost 100 years before Peirce's experiment, Laplace postulated that the frequency of an error could be expressed as an exponential function of the numerical magnitude of the error, disregarding sign \citep{laplace1774}. This is known as Laplace's first law of error.
    \item Gaussian distribution. Laplace proposed Gaussian distribution as the second candidate for the error curve in 1778.
\end{itemize}

These two distributions provide a simple yet believable model-0 to start our search for a realistic error distribution for the Peirce data. Our strategy will be as follows: first, would like to know which of Laplace's laws provides a more reasonable choice as an initial candidate model. In other words, which distribution is relatively \textit{less} surprised by the Peirce data. Second, we like to understand the nature of misspecifications of these two models over the set of $24$ experimental datasets. This will ultimately help us repair $f_0(x)$ by informing us which components are damaged.\footnote{Also, some misspecifications may be harmless as far as the final decision-making is concerned. Knowing the nature of deficiency can help us avoid over-complicating the model-0.}  As John \cite{tukey1969} said: `\textit{Amount, as well as direction, is vital.}' 



\subsection{Informative Component Analysis}

Generating new hypotheses in response to the rejection of the initial candidate model is one of the central objectives of Informative Component Analysis (ICA).

\vskip1em

{\bf Gaussian Error Distribution}.
We devise a graphical \textit{explanatory} method, called Informative Component Analysis (ICA), to perform `informative' data-model comparison in a way that is easily interpretable for large number of parallel experiments like Peirce data. The general process goes as follows:
\begin{center}
Algorithm: Informative Component Analysis (ICA) 
\end{center}
\vspace{-.4em}
\medskip\hrule height .8pt
\vskip1em

\texttt{Step 0.}  Data and notation. For the $t$-th day experiment: we observe ${\bf x}_t = (x_{t1}, x_{t1}, \ldots, x_{tn_t})$ with empirical distribution $\wtF_t$.

\begin{figure}[ ]
  \centering
\includegraphics[width=.65\linewidth,keepaspectratio,trim=2cm 1cm 2cm 1cm]{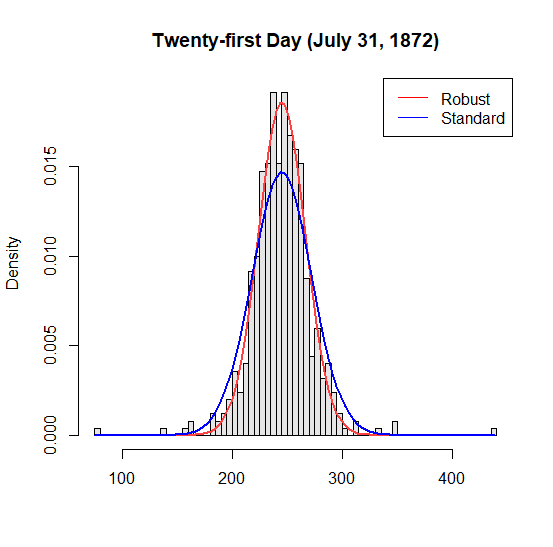}
\vskip.3em
\caption{Two normal distributions are compared with different scale estimates: The red curve is based on the robust IQR-based scale estimate and the blue one is the usual standard deviation-based curve. Clearly, the blue curve underestimates the peak and overestimates the width of the density (due to the presence of few large values in the tails).}\label{fig:rsd}
\end{figure}

\vskip.4em
\texttt{Step 1.}  For each day, we estimate the best-fitted Gaussian distribution $\varphi_t = \cN(\tmu_t, \tsi_t)$, where the parameters are robustly estimated: $\tmu_t$ is estimated by the median and $\tsi_t$ is estimated by dividing the interquartile range (IQR) by $1.349$.
The presence of large outlying observations makes the IQR-based robust-scale estimate more appropriate than the usual standard deviation based estimate of $\sigma_t$; see Fig. \ref{fig:rsd}.  
\vskip.4em

\begin{figure}[ ]
  \centering
\includegraphics[width=.52\linewidth,keepaspectratio,trim=2.2cm 2cm 2.4cm 1.6cm]{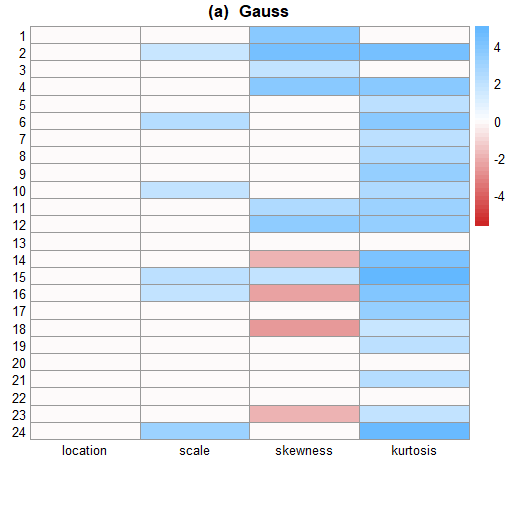}\\[3.5em]
\includegraphics[width=.52\linewidth,keepaspectratio,trim=2.2cm 2cm 2.4cm 1.6cm]{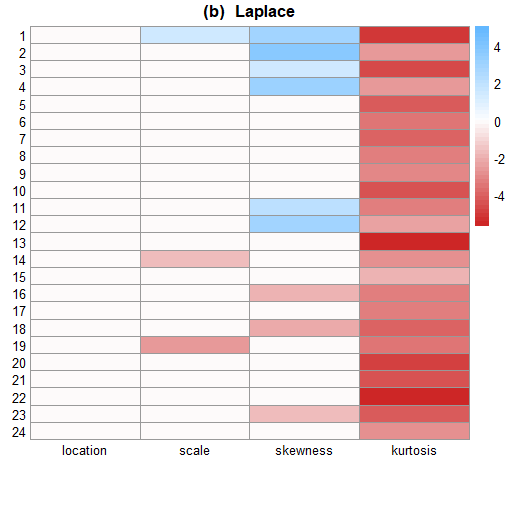}
\vskip1em
\caption{Informative component analysis: LP-Map for Gauss and Laplace models. The rows denote different time points and columns are the order of LP coefficients $\LP[t,j]$. This graphical explanatory method uncovers what are some of the most prominent ways a large number of related distributions differ from the anticipated $f_0(x)$.}\label{fig:lpmapG}
\end{figure}

\texttt{Step 2.} For each experiment, compute the LP-coefficients between the assumed $\varPhi_t$ and the empirical distribution $\wtF_t$.
\beq \LP[j; \Phi_t, \wtF_t]~=~\Ex\big[ T_j(X_t;\Phi_t); \wtF_t \big]~=~\frac{1}{n_t}\sum_{i=1}^{n_t} T_j(x_{ti};\Phi_t).~~~~~\eeq
for $t=1,2,\ldots,24$ and $j=1,\ldots,4$. The smoothed LP-coefficients (applying \texttt{OPEN} model selection method based on AIC-penalty; see equation \ref{eq:open}) are stored in $\LP[t, j]\, :=\, \widehat{\LP}[j; \Phi_t, \wtF_t]$. 
\vskip.4em
\texttt{Step 3.}  LP-Map: Display the $24\times 4$ LP-matrix as an image for easy visualization and interpretation. This is shown in Fig. \ref{fig:lpmapG}(a).
\vskip.5em
\medskip\hrule height .8pt
\vskip1.25em

\vskip.6em
\textit{Interpretation}. What can we learn from the LP-map? It tells us the \textit{nature} of non-Gaussianity of the error distributions for all 24 of the experiments in a compact way. The ICA-diagram detects three major directions of departure (from assumed Gaussian law) that more or less consistently appeared across different days of the experiment: (i) excess variability: This is indicated by the large positive values of the second-order LP-coefficients (2nd column of the LP-matrix) $\{\LP_{t2}\}_{1 \le t\le 24}$. (ii) Asymmetry: It is interesting to note that the values of $\{\LP_{t3}\}_{1 \le t\le 24}$ change from positive to negative somewhere around the 15th day---which implies that the skewness of the distributions switches from left-skewed to right-skewed around the middle of the experiment. (iii) Long-taildness: Large positive values of $\{\LP_{t4}\}_{1 \le t\le 24}$ strongly indicate that the measurement densities are heavily leptokurtic, i.e., they possess larger tails than normal.  These fatter tails generate large (or small) discrepant errors more frequently than expected---as we have witnessed in Fig. \ref{fig:boxplot}.

\vskip.6em

{\bf Laplace Error Distribution}.
Here we choose $f_0(x)$ as the $\texttt{Laplace}(\mu, s)$ distribution:
\[ f_0(x) = {\frac  {1}{2s}}\exp \left(-{\frac  {|x-\mu |}{s}}\right)\,\!, ~~x \in \cR\]
where $\mu \in \cR$ and $s>0$. The unknown parameters are estimated using the maximum likelihood (MLE) method that automatically yields robust estimates: sample median for the location parameter $\mu$ and mean absolute deviation from the median for the scale parameter $s$.

\vskip.3em
The LP-map after applying the ICA algorithm is displayed in Fig. \ref{fig:lpmapG}(b), which shows that a moderate degree of skewness and a major tail-repairing are needed to make Laplace consistent with the data. It is important to be aware of them to build a more realistic model of errors---which is pursued in the next section.

\begin{figure}[ ]
\vspace{-1em}
  \centering
\includegraphics[width=.66\linewidth,keepaspectratio,trim=2cm 1cm 2cm 3cm]{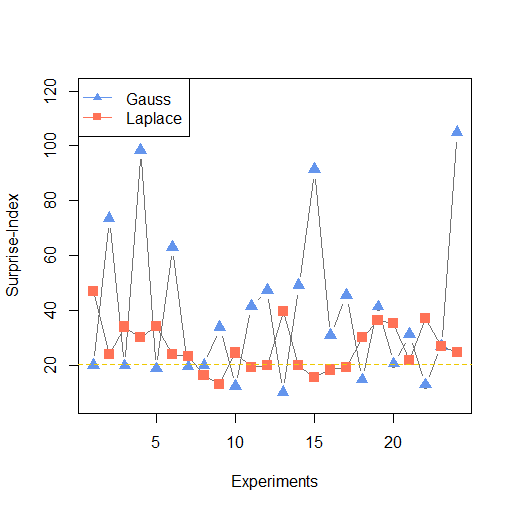}
\vskip1em
\caption{Shows the surprisal-indices \eqref{eq:silp} for the Gauss and Laplace models, over the sequence of $24$ experiments. A `small' value of ${\rm SI}(F_0, \wtF)$ indicates that it is comparatively easier to repair $f_0(x)$ to fit the data. The plot provides a clear rational basis for choosing Laplace as a preferred model-0 since the blue curve consistently exceeds the orange curve for most experiments.
}\label{fig:Glapinfor}
\end{figure}

\vskip.5em
{\bf Laplace or Gauss?} Following \eqref{eq:sindex}, compute
\beq \label{eq:silp}
{\rm SI}(F_{0t}, \wtF_t)~=~\sum_{j=1}^4\Big |\, \hLP[t,j]\, \Big|^2,\eeq
by taking the sum of squares of each row of the LP-matrix.  Fig. \ref{fig:Glapinfor} compares the surprisal-index for the normal and Laplace distribution over $24$ experiments. From the plot, it is evident that Peirce's data were better represented by Laplace than by Gaussian. 

\vspace{-.65em}
\subsection{Examples} \label{sec:ex}
The primary goal here is to show how density-sharpening provides a \textit{statistical method for repairing a scientifically meaningful model} based on observed data. To that end, we apply the theory of density-sharpening to two specific day studies with $f_0(x)$ as the Laplace distribution. 


\begin{figure}[ ]
\vspace{-1.5em}
  \centering
\includegraphics[width=.38\linewidth,keepaspectratio,trim=1cm 1cm 1cm 1cm]{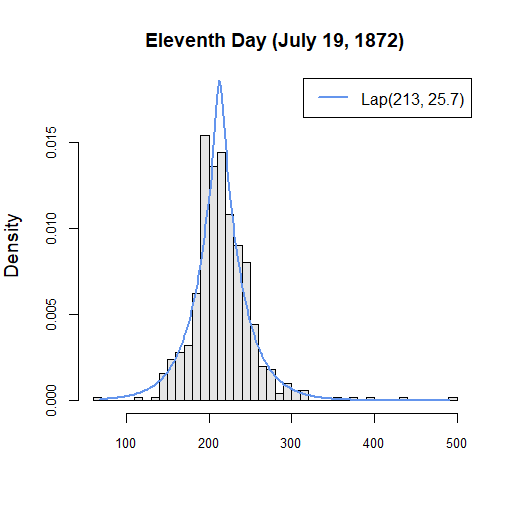}~~~~~~
\includegraphics[width=.38\linewidth,keepaspectratio,trim=1cm 1cm 1cm 1cm]{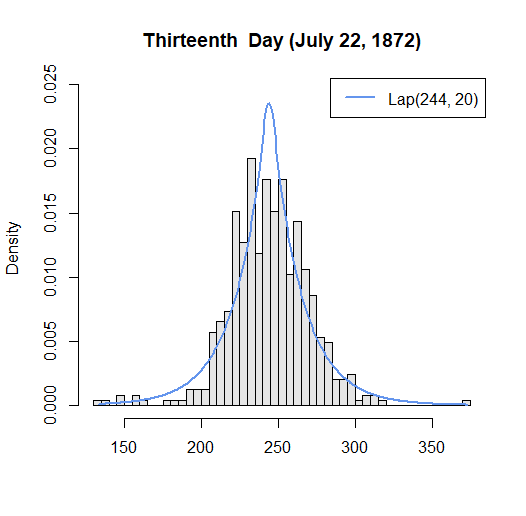}\\[.34em]
\includegraphics[width=.38\linewidth,keepaspectratio,trim=1cm 1cm 1cm 1cm]{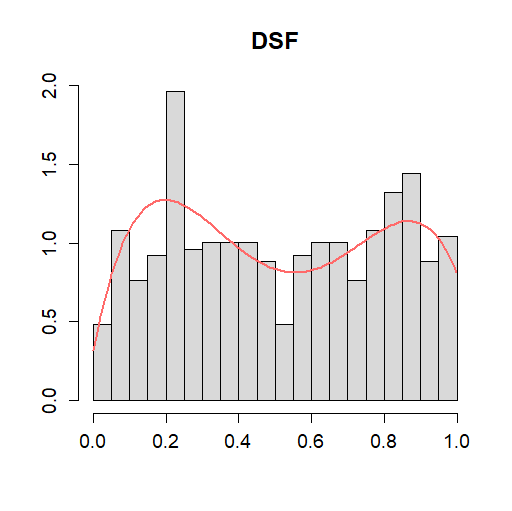}~~~~~~
\includegraphics[width=.38\linewidth,keepaspectratio,trim=1cm 1cm 1cm 1cm]{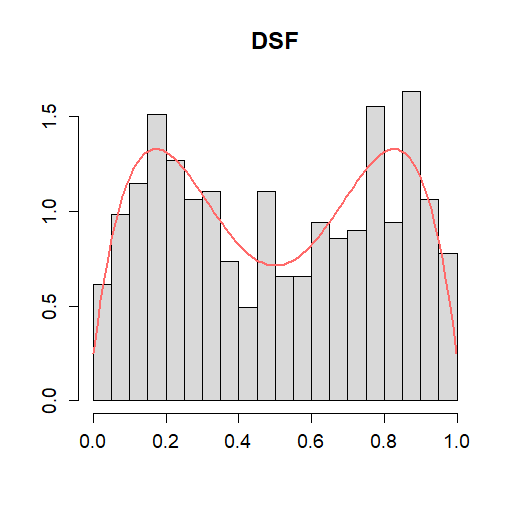}\\[.34em]
\includegraphics[width=.38\linewidth,keepaspectratio,trim=1cm 1cm 1cm 1cm]{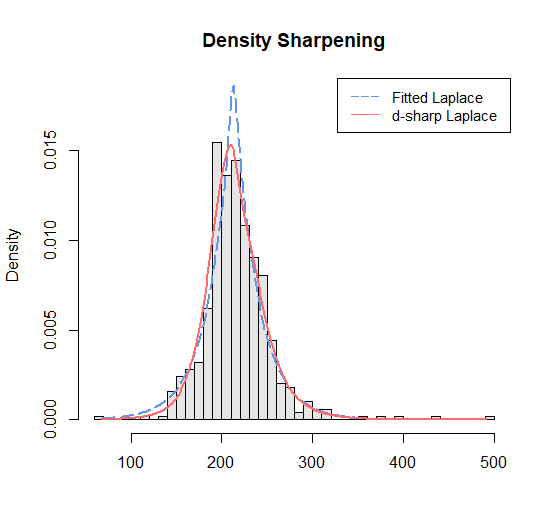}~~~~~~
\includegraphics[width=.38\linewidth,keepaspectratio,trim=1cm 1cm 1cm 1cm]{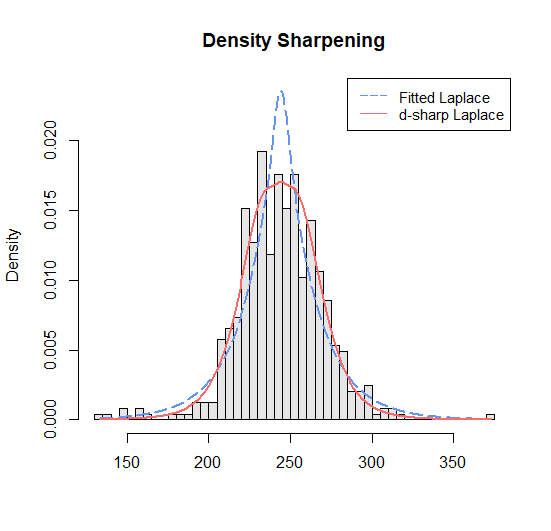}
\vspace{-.25em}
\caption{Mechanism of density-sharpening for 11th (1st column) and 13th (2nd column) day experiments. Top row: Display the best-fitted Laplace distributions $f_0$. Middle row: Displays the estimate $\whd(u;F_0,\wtF)$, which reveals the nature of statistical uncertainties of the Laplace models. Put it simply, the shape of $\whd(u;F_0,\wtF)$ answers the central question of discovery: What have we learned from the data that we did not already know? This helps to \textit{invent} new hypotheses that are worthy of pursuit. Last row: Estimated \texttt{SharpLaplace} models $f_0(x) \times \whd_0(x)$ are shown (red curves). Here $\whd_0(x)$ rectifies the shortcomings of Laplace model by ``bending'' it in a data-adaptive manner. }\label{fig:sharplap}
\end{figure}

\vskip.7em
{\bf Study \#11 (July 19, 1872).} The best fitted Laplace distribution with mean $213$ and scale parameter $25.7$ is shown in the top left of Fig. \ref{fig:sharplap}. The estimated sharpening kernel with smoothed LP-coefficients (see Sec. \ref{sec:rtheory}) is given below: 
\beq 
\label{eq:exp11d}
\setlength{\abovedisplayskip}{1em}
\setlength{\belowdisplayskip}{1em}
\whd_0(x)\, :=\, \whd(F_0(x); F_0, F)\,=\, 1 + 0.095 T_3(x;F_0) -0.148 T_4(x;F_0),
\eeq
and is shown in the middle panel.

The graphical display of $d_0(x)$ provides actionable insights into \textit{how} to modify the Laplace distribution to reduce the empirical surprise. The non-zero $\hLP_3$ and $\hLP_4$ indicates that the Laplace distribution needs to be corrected for skewness and kurtosis, which is accomplished via LP-orthogonal series $\DS(F_0,m)$ model \eqref{DSm1}:
\beq \label{eq:s11}
\hf(x)~=~{\frac  {1}{2s_0}}\exp \left(-{\frac  {|x-\mu_0 |}{s_0}}\right)\,\Big[ 1\,+\,0.095 T_3(x;F_0)\, -\,0.148 T_4(x;F_0)\Big],
\eeq
where $(\mu_0,s_0)=(213, 25.7)$. The model 
\eqref{eq:s11} sharpens the assumed Laplace law to render it more closer to the observed fact. The bottom-left panel of Fig.  \ref{fig:sharplap} displays the asymmetric Laplace distribution with a shorter left-tail. The maxent $\DS(F_0,m)$ estimate
\beq \label{eq:s11exp}
\hhf(x)~=~{\frac  {1}{2s_0}}\exp \left(-{\frac  {|x-\mu_0 |}{s_0}}\right)\,\exp\Big \{ \,0.098 T_3(x;F_0)\, -\,0.153 T_4(x;F_0)\,-\,0.0152 \Big\},
\eeq
whose shape is almost indistinguishable from that of \eqref{eq:s11}, and thus not displayed.

\vskip.64em
{\bf Study \#13 (July 22, 1872).} We apply the same steps to derive the error distributions of the day-13 experiment. We choose $f_0(x)$ as Laplace$(244, 20)$ and estimate the density-sharpening function:
\vspace{-.1em}
\beq 
\whd_0(x)\, :=\, \whd(F_0(x); F_0, F)\,=\, 1  -0.256 T_4(x;F_0).
\eeq
The shape of $\whd_0(x)$ clearly indicates that the peak and the tails of the initial Laplace distribution need repairing.  

Ampliative character of our modeling: The DSF $\whd_0(x)$ allows the Laplace to identify its own limitations and drives it to evolve into a new, more complete one:

\beq \label{eq:s13}
\hf(x)~=~{\frac  {1}{2s_0}}\exp \left(-{\frac  {|x-\mu_0 |}{s_0}}\right)\,\Big[ 1\,-0.256 T_4(x;F_0)\Big],
\eeq
with $(\mu_0,s_0)=(244, 20)$; shown in the bottom-right panel of Fig.  \ref{fig:sharplap}. Compared with the Laplace distribution (the blue curve), the $d$-modified Laplace (the red curve) is much wider with a rounded peak and clipped tails.




\section{A Generalized Law of Errors}

We have seen that for Peirce data, Laplace distribution was surprised in different manners for different experiments (refer Fig. \ref{fig:lpmapG}b and Fig. \ref{fig:sharplap}): e.g., on day 11, the Laplace model got puzzled by the discrepancy in skewness, and tail of the measurement distribution, whereas on day 13, the surprise was mainly due to tail differences. The question naturally arises: how should a Laplace model respond to unexpected changes in data? A proposal for generalized law of errors is given that allows Laplace to automatically adapt to new data environments.

\vskip1em
\begin{defn}[Self-improving Laplace model]
We call $X \sim $ \texttt{SharpLaplace}($m$), when the density of $X$ obeys the following parameterizable form:
\beq 
\label{weq:slap}
f(x)~=~{\frac  {1}{2s}}\exp \left(-{\frac  {|x-\mu |}{s}}\right)\,\exp\Big \{ \sum_{j=1}^m \te_j T_j(x;F_0)\,-\, \Psi(\teb)\Big\},~~~x \in \cR.~~~
\eeq
The insights gained from the analysis done in Sec. \ref{sec:ex} suggest that $m=4$ or $6$ 
could be perfectly reasonable for most practical purposes. The power of this model lies in its capacity to self-modify its structure in a data-driven manner.
\end{defn}


\begin{rem}
\texttt{SharpLaplace} class of error models has inbuilt ``rules'' (principles and mechanisms) that tell a Laplace how to adapt with the data in a completely autonomous fashion without being pre-programmed into them. This auto-adaptive nature makes this model realistic enough to be useful for a wide range of scientific applications.
\end{rem}

\section{Peirce's Law of Discovery}
\begin{quote}
    \textit{Not the smallest advance can be made in knowledge beyond the stage of vacant staring, without making an abduction at every step. \hfill  {\rm --- C. S. \cite{peirce1901proper}}}   
\end{quote}


All empirical scientific inquiry goes through three fundamental inferential phases:
\begin{itemize}[itemsep=1pt,topsep=1pt]
 \setlength{\itemindent}{.7em}
    \item Discovery: developing new testworthy hypotheses;
    \item Hypothesis testing: confirming the plausibility of a hypothesis;
    \item Prediction: predicting by extrapolating the acceptable model.
\end{itemize}

Over the last century or so, statistical inference has been dominated by hypothesis testing and prediction\footnote{Discovery is much harder than prediction because one can go away with good prediction without understanding. But for discovery, understanding `how and why' is a must.} problems, virtually neglecting the key question of where the reasonable hypothesis came from, leaving it to the scientists'  imagination and speculation. In the article `Statistics for Discovery,' George E. P. Box argued that
\begin{quote}
   \textit{[S]tatistics has been overly influenced by mathematical methods rather than the scientific method and consequently the subject has been greatly skewed towards testing rather than discovery.} \hfill {\rm --- George \cite{box2001dis}}
\end{quote}

Our focal interest is in the problem of discovery, not confirmation or prediction. We showed how density-sharpening based modeling can provide the basis for developing statistical laws of discovery.


\vskip.6em

Charles Sanders Peirce introduced the idea of \textit{abductive inference}  (as opposed to inductive inference) to describe the process of generating hypotheses in order to explain surprising facts. He developed abductive reasoning over $50$ years (between 1865 and 1914), and it is considered as Peirce's most significant contribution to the logic of science. According to Peirce, abduction `is the only logical operation which introduces any new idea.'\nocite{peirce1931collected}
The importance of abduction for scientific discovery was further stressed by \cite{heckman2017abducting}:  
\vspace{-.4em}
\begin{quote}
{\small \textit{Abduction is different from falsification or corroboration. It moves descriptions of the world forward rather than just confirming or falsifying hypotheses. It is part of a process of discovery where model reformulation, revision of hypotheses and addition of new information are part of the process.}}
\end{quote}

\vspace{-.4em}







\vskip.6em

{\bf Abductive Inference Machine (AIM)}. 
This paper lays out a proposal for algorithmic operationalization
of Peircean style abductive inference and discovery. In particular,  we described how the density-sharpening principle can help us design an Abductive Inference Machine (\texttt{AIM}\footnote{It tells empirical analysts where to AIM as they search for possible new discoveries.}) that (i) allows us to properly \textit{handle} model uncertainty and misspecifications; (ii) produces \textit{abductive instinct}---by guiding us to make better decisions (than depending on pure luck alone) in formulating and adopting new promising hypotheses that have a better chance of being true; and (iii) generates a preferred course of \textit{actions} for extracting statistical models from experimental data by revising an initially misspecified scientific model.


\section{AIM: Science of Model Development and Revision}
\label{sec:AIMsec}
\texttt{AIM} is a theory of \textit{model-revision}, not parameter estimation (MLE/Bayes/robust methods) or curve-fitting (machine learning methods). There are some unique objectives and challenges, which set it apart from traditional data modeling cultures. In the following, we will highlight a few major ones (D1-D7), taking help from the Peirce data analysis done in Section \ref{sec:analysis}.





\vskip.4em
Four stages of abductive model building:
\vskip.2em
~~1.~{\bf Initial state.} \texttt{AIM} starts with an approximate model $f_0$ (based on, say, some economic theory) and measurements.\footnote{In our context, the theory of Laplace's law of error was confronted with Peirce's experimental data.} The top left density in Fig. \ref{fig:sharplap} shows the best-fitted theoretical model---the Laplace distribution for experiment \#11, where the unknown parameters were estimated using MLE.
\vskip.52em

\circled{D1} \textit{Non-standard inferential questions: Justification $\rightarrow$ discovery}.  However, for modern econometricians and policymakers, parameter estimation or significance testing routines (classical inference; see \cite{haavelmo1944prob}) are not the most interesting issues. Modern quantitative economists are more concerned with questions like: 
``How far is our speculated model from reality? What are the most important gaps in our understanding? In which directions can I improve my theory-based model?''  Developing a general approach to answering these modeling questions is the central imperative of \texttt{AIM} .

\vskip.52em
~~2.~{\bf Encountering surprise}. `Surprise' jumpstarts the abductive learning process. Surprise essentially means that there is something new in the data \textit{relative} to the assumed model $f_0$, which we estimate by ${\rm SI}(F_0, F)$ following Eq. \eqref{eq:sindex}. Simply put, it quantifies how much \textit{new} information is left on the data to be explained; see the orange curve in Fig. \ref{fig:Glapinfor}. Intelligent learners (agents) utilize surprise as a source of additional information to learn something new about the phenomena.




\vskip.5em
\circled{D2} \textit{Model-disequilibrium theory}. A `large' value of ${\rm SI}(F_0, F)$  indicates the model is `out of equilibrium' with the current environment, and to restore equilibrium, a careful revision of the current theoretical model (beyond parameter tuning) is necessary. But how do we sharpen the existing model? Can we develop an \textit{automatic} procedure to generate the sharpening rules? These questions are beyond the reach of classical statistical learning methods. \texttt{AIM} approaches these questions by first characterizing the ``knowledge-gap'' between the postulated theory and the observed measurements.






\vskip.5em
~~3.~{\bf Discovering the knowledge-gap}. Fig. \ref{fig:sharplap} (left of middle panel) displays the estimated sharpening kernel $\whd(u;F_0,F)$ for experiment \#11, which acts as a `channel' through which information flows from the data to the model---obeying the density-sharpening principle---to bring the \textit{system} (model-data) back to equilibrium. Accordingly, $\whd(u;F_0,F)$ acts as a  ``bridge'' between a theorized  model and actual measurements. The following remark by Trygve \cite{haavelmo1944prob} highlights how crucial this accomplishment is:
\vspace{-.4em}
\begin{quote}
{\small \textit{The method of econometric research aims, essentially, at a conjunction of economic theory and actual measurements, using the theory and technique of statistical inference as a bridge pier.}}
\end{quote}

\vskip.52em
\nocite{hanson1965patterns}
\circled{D3} \textit{Hypotheses generation}. Abduction is the logic of discovery. Why do we need a \textit{logic} for discovery? Charles Peirce, Herbert Simon, and many other prolific researchers believed that a \textit{trial-and-error} search for ``invention'' is seldom a worthwhile strategy, especially for complex systems (like economics, biology, etc.). Norwood Hanson said it beautifully in his book:
\vspace{-.3em}
\begin{quote}
    ``\textit{If establishing an hypothesis through its predictions
has a logic, so has the conceiving of an hypothesis.}'' \hfill---Patterns of Discovery (1958)
\end{quote}

\texttt{AIM} helps scientists to make educated guesses---on what’s the next best hypothesis to try from a vast pool of conceivable collections---by autonomously learning new realities from the data through $d_0(x)$. Classical inference, on the other hand, mainly deals with the confirmatory or predictive side of data analysis, not hypotheses generation and discovery. 

\vskip.52em
\circled{D4} \textit{Modeling surprise, not the full data}. 
\texttt{AIM} searches for patterns in the ``unexplained rest''---the parts of the data that were left \textit{unexplained} by the existing theory. Notice that we are not blindly searching for patterns in the full data; we are only focusing on the \textit{novel parts} of the data that contain new information. It is important to distinguish between these two aspects. Our goal is to accelerate discovery by synthesizing a simple explanatory model $\whd(u;F_0,F)$ for the surprising phenomena. 

\vskip.52em
\circled{D5} \textit{Information-filtering unit}.
The density-sharpening kernel acts as a filter that discards redundant information and compactly summarizes the new information (knowledge-gap) as a probability density function.\footnote{As Herbert Simon said: ``Anything that gives us \textit{new knowledge} gives us an opportunity to be more rational.'' From that perspective, \texttt{AIM} could be a powerful tool to guide economic agents in making rational decisions under uncertainty. More details can be found in \cite{D22ADM}.} The non-zero LP-coefficients of $\whd(u;F_0,F)$ identify the missing elements of reality in the current theoretical model. For example, Eq. \ref{eq:exp11d} implies that the Laplace law is misspecified in terms of symmetry (3rd order) and long-tailedness (4th order); also check the 11th row of the LP-map shown in Fig. \ref{fig:lpmapG}. Standard statistical learning methods don't have any such capabilities.










~~4.~{\bf Model-editing}.  \texttt{AIM} fills the knowledge-gaps by revising initial probability model based on the principle of density-sharpening, which can be described by a simple logical formula: 
\beq 
\setlength{\abovedisplayskip}{1.2em}
\setlength{\belowdisplayskip}{1em}
\label{eq:leq}
\fbox{\text{Hypothesized-Model}} ~~+ ~~ \fbox{\text{New data}}~~~~\xRightarrow[\texttt{Sharpening}]{d(u;F_0,F)} ~ ~~~\fbox{\text{Improved-Model}}~~~~
\eeq

Every successive iteration of the above procedure generates a more realistic model than its predecessors. Here the density-sharpening kernel $d_0(x)$ represents the \textit{progress} in our understanding, which makes the hypothesized simplified model elastic enough to be adaptable for real-data.\footnote{In other words, we don't believe in the `one-fits-all' model. Our goal is to provide economists with a systemic principle for iteratively revising their preliminary models by confronting them with real-world data.}  

By executing \eqref{eq:leq}, \texttt{AIM}  \textit{designs} a class of most pursuitworthy alternative models for the data and \textit{selects} the best one using the `law of lazy update.'\footnote{It is fundamentally different from model selection or multiple hypothesis testing, which deals with a pre-determined set of alternative models. Model discovery and model selection are two very different things.} Eq. \ref{eq:s11exp} shows the sharp-Laplace model for study \#11, also displayed in the bottom of Fig. \ref{fig:sharplap}. The parsimonious $d_0(x)$ keeps the final model `sophisticatedly simple'\footnote{Arnold \cite{zellner2007}: `a much heavier emphasis on sophisticated simplicity in econometrics is needed.'} by \textit{smoothly} extending the hypothesized Laplace model to explain the data.


\vskip.52em
\circled{D6} \textit{AIM $\ne$ Curve-fitting}. 
One of the non-standard aspects of \texttt{AIM} is that it's not about building fancy empirical models starting from a clean slate---it's about building a statistical structure \textit{on top of} the already existing scientific knowledge base to advance the current theory.\footnote{In more simple terms, \texttt{AIM} = Learning from data by standing on the foundation of existing knowledge.} In doing so, it provides efficient ways of handling an idealized simple model for discovering \textit{new knowledge} from complex real-world data.

\vskip.64em
\circled{D7} \textit{Conservative-liberal coalition}. Note that the derived sharp-probabilistic law (e.g., eq. \ref{eq:s11exp}) combines the \textit{generality} (generic features) of Laplace's laws of error with the \textit{specificity} (stylized features) of Peirce's data. This coalition of conservative (sticking to current dogma) and liberal (openness to course correction when necessary) ideologies makes our data modeling philosophy stands out from traditional statistical and machine learning data-fitting methods.    

\section{AI = Abductive Intelligence}
When can we say a model is behaving intelligently?  This is no easy question. However, at least part of the answer, I believe, lies in inspecting \textit{how} the model reacts to surprise and adapts to changes. An intelligent model should not be `brittle,' which collapses all of a sudden upon encountering surprises from the data.  The model should have an `internal' mechanism that help it rise to the occasion by providing recommendations for how to put together incomplete pre-existing knowledge (encoded in model-0) with the data. We refer to this core intellectual component of \textit{any} learning problem as ``abductive intelligence.'' 






\vskip.55em


Building models that are capable of improving themselves has been a dream of computer scientists since the inception of the artificial intelligence field in 1956. It was at the top of the agenda in the proposal written for the Dartmouth Summer Research Project on artificial intelligence:


\begin{quote}
    ``\textit{Probably a truly intelligent machine will carry out activities which may best be described as self-improvement.}'' \hfill---John \cite{mccarthy55}
\end{quote}

More than developing new ways of building data models, we need new principles for sharpening an existing model's infrastructure. Such a model, equipped with self-improving capability, then gradually acquires more knowledge about the environment by building increasingly refined models of reality. 
\begin{quote}
    \textit{``Once we have devised programs with a genuine capacity for self-improvement a rapid evolutionary process will begin. As the machine improves both itself and its model of itself, we shall begin to see all the phenomena associated with the terms `consciousness,' `intuition' and `intelligence' itself.''} \hfill ---Marvin \cite{minsk66}
\end{quote}

\begin{rem}[Intelligence of a Model]
A model's `intelligence' is its capacity to change and remodel itself when confronted with new data. Any model which is not capable of `self-improving' is a disposable, dead model. 
\end{rem}

 \begin{rem}[Designing an Intelligent Model]
 Model is a work in progress; there is no such thing as the `final model.' The important part is knowing \textit{how} to expand the knowledge base by sharpening yesterday's version. With that being said, the focus of present-day machine learning has primarily been on developing a good subordinate-model $f_0(x)$ but not so much on designing the supervisory-model $d_0(x)$. To achieve the goal of designing intelligent machines, it seems inevitable that we have to shift our attention to the `supervisory' part of the model---that which knows how to react to surprise in order to evolve to its next avatar. In the long run, the durability of a model depends more on its \textit{skill to adapt} than its  \textit{built-in skill}. And to design such  self-improving models, we will need to instill `abductive intelligence' in the existing machine learning systems---which is the goal of \texttt{AIM}.  
 \end{rem}




\section{Peirce: An Explorer Upon Untrodden Ground}
\vspace{1em}
\begin{center}
  \hspace{1.3in}  \textit{I was an explorer upon untrodden ground}    \hfill --- C. S. \cite{peirce1902minute}
\end{center}

Bertrand \cite{russell1959wisdom} described  Charles Sanders Peirce as ``one of the most original minds of the later nineteenth century, and certainly the greatest American thinker ever.''  He was a craftsman of the highest order, who made some eminent  contributions to the development of nineteenth-century American Statistics. Peirce analyzed his 1872 experimental data in the paper ``On the theory of errors of observation''---which is a gold mine of ideas. His techniques and philosophy of data analysis reported in the paper were a testament to his brilliance as a master applied statistician. On the landmark occasion of 
150th anniversary of his famous 1872 study,\footnote{A miraculous year: In 1872, exactly the same year of Peirce's experiment, Ludwig Boltzmann established the probabilistic (statistical) foundation of the \textit{entropy} function---the birth of modern information theory. As we have seen in this article, Peircean abduction and  information theory are intimately connected through the concept of density-sharpening. As a matter of fact, \texttt{AIM} stands on three pillars: Abductive inference + information theory + density-sharpening law.} we have looked back at his views on empirical modeling. The current article offers a framework that embraces and operationalizes the Peircean view of discovery and statistical modeling. We called this framework \texttt{AIM}--Abductive Inference Machine---which is grounded in the principle of density sharpening and a new class of models called ``dyadic models.'' We have illustrated the key algorithmic steps and philosophical aspects
of our modeling approach using Peirce's 1872 experimental data to reveal new insights on the probabilistic nature of measurement uncertainties.

\section*{Statements and Declarations}
The Author declares that there is no conflict of interest/competing interests.

\section*{Acknowledgement}
The author was benefited from conversations and/or correspondence with James Heckman, Burton Singer, and Stephen Stigler. We also thank the editorial board for their thoughtful and philosophical comments on the manuscript.

\vspace{.6em}

2022 marks the 150th anniversary of C. S. Peirce’s historic experiment---which according to Stigler (1978) is one of the most significant statistical investigations in the history of nineteenth-century American mathematical research. This paper celebrates this landmark occasion by introducing a modern approach to Peircean abduction, which is readily applicable in practice.

\bibliographystyle{Chicago}
\bibliography{ref-bib}

\end{document}